\documentstyle[twoside,fleqn,espcrc2,psfig]{article}

\newcommand{\be}{\begin{equation}}
\newcommand{\ee}{\end{equation}}
\newcommand{\bea}{\begin{eqnarray}}
\newcommand{\eea}{\end{eqnarray}}
\newcommand{\bes}{\begin{eqnarray*}}
\newcommand{\ees}{\end{eqnarray*}}

\newcommand{\AmS}{{\protect\the\textfont2
   A\kern-.1667em\lower.5ex\hbox{M}\kern-.125emS}}

\newcommand{\sint}{\sin^2{\theta_W}}

\title{Electroweak Unification into a Five-Dimensional SU(3) at a TeV}

\author{Savas Dimopoulos\address{Physics Department, Stanford
University, Stanford, CA 94305-4060, USA.}, David
Elazzar Kaplan\address{SLAC, Stanford, CA 94025} and Neal
Weiner\address{Department of Physics, University of Washington,
Seattle, WA  98195}
          }

\begin{document}

\begin{abstract}

We apply a recently proposed mechanism for predicting the weak
mixing angle to theories with TeV-size dimensions.
``Reconstruction'' of the associated moose (or quiver) leads to
theories which unify the electroweak forces into a five
dimensional SU(3) symmetry. Quarks live at an orbifold fixed point
where $SU(3)$ breaks to the electroweak group. A variety of
theories -- all sharing the same successful prediction of
$\sint$-- emerges; they differ primarily by the spatial location
of the leptons and the  absence or presence of supersymmetry. A
particularly interesting theory puts leptons in a
Konopinski-Mahmoud triplet and suppresses proton decay by placing
quarks and leptons on opposite fixed points.

\vspace{1pc}
\end{abstract}

\maketitle
\section{Introduction}
The one quantitative success of physics beyond the standard model (SM)
is the prediction of the weak mixing angle $\sint$ by supersymmetric
grand unified theories (GUTs).  The $SU(5)$ prediction is $\sint=3/8$
at tree level \cite{Georgi:1974sy}.  Running this value from the GUT scale to
the weak scale in supersymmetric theories produces the measured 
value of $0.231$ within theoretical uncertainties \cite{Dimopoulos:1981zb}. 
A critical assumption 
is the existence of a large energy desert above the weak scale.  For a 
theory with a low cutoff, an alternative approach must be taken.

Recently, a new mechanism for predicting the weak mixing
angle with TeV-physics was proposed \cite{Dimopoulos:2002mv}. It leads to
the unification of the two electroweak gauge couplings into
their $SU(3)$-symmetric value, giving $\sint=.25$ at tree level. Since
this value is close to the experimental value of
$\sint=.231$ at $M_Z$, SU(3)-unification occurs at a few
TeV \cite{Dimopoulos:2002mv}. The proposed mechanism is quite simple: 
one adds to the SM an $SU(3)$-gauge
group and a scalar $\Sigma$ whose vacuum expectation value
(VEV) breaks the $SU(3)$ and the electroweak gauge sector
to the diagonal $SU(2)\times U(1)$. SM fields are all
singlets under the new $SU(3)$. Nevertheless, if the
original $SU(2)\times U(1)$ couplings are somewhat large,
the low energy gauge groups reflect the $SU(3)$ symmetry.

Recent studies of ``dimensional deconstruction'' 
\cite{Arkani-Hamed:2001ca,Hill:2000mu} have
highlighted the close connection between ``moose'' or ``quiver'' theories
and those with extra dimensions. With this motivation, we will
attempt to ``reconstruct'' the fundamental moose
of this mechanism (figure 1a) into its one-dimensional
cousin. The models which result (figure 1b)
contain a five-dimensional $SU(3)$ broken to $SU(2)\times
U(1)$ at one orbifold fixed point on which some or all the SM
particles are localized. The purpose of this paper is to
study the physics of the simplest model and its variations.

In these models quarks must be localized on the
$SU(2)\times U(1)$ orbifold fixed point. If quarks lived in
the bulk or the $SU(3)$-symmetric fixed point, they would
have to belong to $SU(3)$ multiplets, and this would
conflict with their fractional charges \cite{Weinberg:1972nd}.
On the other hand, integrally charged particles -- the
leptons and the Higgs-- can belong to $SU(3)$-multiplets
and therefore can live at either fixed-point or in the
bulk. In fact the lepton doublet and singlet together
neatly fit into a Konopinski-Mahmoud triplet 
\cite{Konopinski:1953gq}. Migrating the
lepton or the Higgs to different locations gives rise to
different theoretical possibilities with distinct
phenomenology.

Another degree of freedom which leads to different
theoretical options is the presence or absence of
supersymmetry. The unification scale of these theories is
in the multi-TeV range. For a Higgs mass is
$\sim 100$ GeV, there is typically a moderate fine tuning
of order of $\sim 10^{-4}$. One way to avoid it is
supersymmetry.
\vskip 0.15in

{\bf \noindent Orbifolds versus Mooses:}

While the minimal module proposed in reference
\cite{Dimopoulos:2002mv} is quite simple, it is interesting to
study five-dimensional realizations of the mechanism.
A reason for this is the appeal of geometric intuition.  For
example, the mechanism of \cite{Dimopoulos:2002mv} requires that
the $SU(3)$ gauge coupling is smaller than that of the original
$SU(2)\times U(1)$. Placing $SU(3)$ in a moderately large bulk
naturally leads to a Gaussian dilution of its coupling strength.
Also, the presence of the bi-fundamental field $\Sigma$ may appear
less natural than breaking the symmetry by orbifold boundary
conditions. Another tool is locality, which can help suppress
proton decay and other dangerous operators \cite{Arkani-Hamed:1999dc}. 
Aesthetic
arguments aside, the important feature of the mechanism proposed
in \cite{Dimopoulos:2002mv} is that it is simple and the
prediction of $\sint$ so robust that it can can be embedded in a
variety of frameworks, including TeV-dimensions.

In section 2 we discuss the prediction of the weak mixing
angle and the associated theoretical uncertainties. This
discussion parallels the one in reference \cite{Dimopoulos:2002mv} and
establishes the correspondence between the deconstructed
and reconstructed minimal modules. In section 3 we analyze
the phenomenology of the minimal reconstructed module and
some of its variations, which differ from each other by the
location of the lepton and the Higgs. An interesting
variation has automatic proton stability due to the
separation of quarks and leptons at opposite fixed points
\cite{Arkani-Hamed:1999dc}. In section 4 we discuss models which address the
hierarchy problem via supersymmetry, and we conclude in
section 5.

\section{Framework}
Before embarking on discussions of specific models, there are a number of
issues which need to be addressed. Specifically: in what sense can a gauge 
theory broken by orbifolding be considered a unified theory? 
How do gauge couplings run in 
five dimensions and can we still discuss ``logarithmic'' evolution of gauge 
couplings? What are the uncertainties and under what circumstances can one 
make quantitative predictions? All of these issues have been discussed in
numerous previous works \cite{Nomura:2001mf,Hall:2001pg,Barbieri:2001cz,Hebecker:2001wq,Hebecker:2001jb,Contino:2001si}, and we summarize here merely the 
essential results.

\begin{figure}
  \centerline{
    \psfig{file=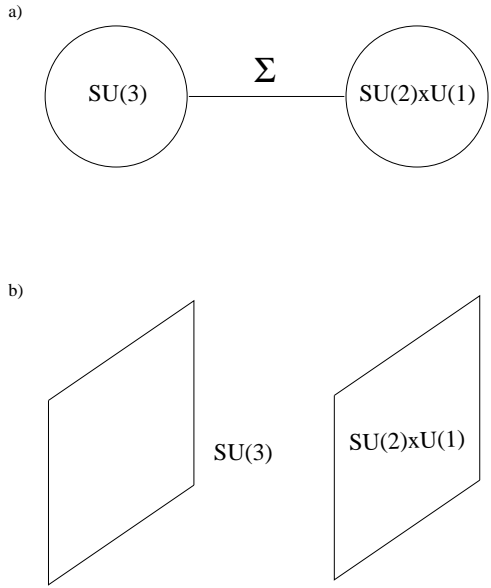,width=0.4\textwidth,angle=0}}
        \caption[sniff]{a) Deconstructed (4D) and b) reconstructed (5D) 
versions of electroweak $SU(3)$ models.}
\label{drcon}
\end{figure}

\subsection{Orbifold Breaking of Symmetries}
We use the simplest geometric orbifold which compactifies a flat
extra dimension, namely ${\cal R}/({\cal Z}_2 \times {\cal Z}_2')$.
Symmetries of the field theory can be broken by projecting out states
which are even or odd under either ${\cal Z}_2$.  The only requirement
is that the projection is consistent with any ${\cal Z}_2$ ``parity'' 
symmetry which exists in the Lagrangian, {\it e.g.}, all even modes of 
negative parity states are projected out.  This operation can break
gauge symmetries, flavor symmetries and supersymmetries and 
is generically referred to as the Scherk-Schwarz mechanism \cite{Scherk:1979ta,Scherk:1979zr}. Within the context of gauge theories, it is also referred to as the Hosotani mechanism \cite{Hosotani:1983xw}.

Within the specific example of an $SU(3)$ theory broken by orbifold 
boundary conditions to $SU(2)\times U(1)$, the gauge fields corresponding 
to the generators of $SU(3)/(SU(2)\times U(1))$ are given odd boundary 
conditions (parity $-1$) about a boundary at $y= \pi R$. Because of this, 
while we can perform arbitrary $SU(3)$ gauge transformations in the bulk, on 
the brane only $SU(2)\times U(1)$ transformations are non-trivial. Due to 
this less restrictive symmetry we can include incomplete $SU(3)$ multiplets 
on the boundary, without any inconsistency in the theory. 
This will allow us to include quarks in our models although they cannot
be put into $SU(3)$ multiplets.  We will also use orbifolding to break
supersymmetry when relevant.

\subsection{Theoretical Uncertainties}
Because $SU(3)$ is broken at the $y= \pi R$ boundary,
we can include contributions to the kinetic terms of the 
gauge fields on the boundary which are only $SU(2)\times U(1)$ 
invariant.
\bea
{\mathcal{L}} &=& \int d^4 x \> dy \frac{F_{ij}^{3} F^{ij}_{3}}{4 g_3^2}+
\gamma F_{\mu \nu}^{3} F^{\mu \nu}_{3}\delta(y) \\&+& (\frac{\delta_Y}{3} 
F_{\mu \nu}^1 F^{\mu \nu}_1 +\delta_2 F_{\mu \nu}^2 
F^{\mu \nu}_2)\delta(y-\pi R).
\eea
Here $i,j$ and $\mu, \nu$ are five and four dimensional Lorentz indices, 
respectively, while $3,2,1$ index whether it is the field strength tensor 
for the complete $SU(3)$ multiplet, or just the $SU(2)$ or $U(1)$ 
subgroup. Note that $g_5^2$ has dimensions here of mass$^{-1}$.

Ignoring quantum effects, we match to the effective four-dimensional 
gauge theory. The gauge couplings are given by
\bea
\frac{1}{g'^2} &=& \frac{3 \pi R}{g_3^2}+3\gamma + \delta_Y,\\
\frac{1}{g^2} &=& \frac{\pi R}{g_3^2} +\gamma+ \delta_2.
\nonumber
\eea
and thus the degree to which the $SU(3)$ relation $g^2/g'^2=3$ depends 
crucially on the size of $\delta_Y$ and $\delta_2$. The $SU(3)$ universal 
piece $\gamma$ is irrelevant for this and we will henceforth ignore it.  
We can gain insight into their size by studying the strength of coupling 
of the theory at the cutoff scale $\Lambda$. That is, we naturally expect
\be
\frac{1}{\Lambda g_5^2} \sim \delta_i
\label{error}
\ee
It is straightforward to make a connection with the uncertainties discussed in \cite{Dimopoulos:2002mv}. In that case, there were uncertainties due to $\tilde g_{1,2}$, which, in the case of strong coupling, perturbed the prediction for $\sint$ only slightly. Here, the equivalent uncertainties arise from $\delta{Y,2}$, which are small when the theory is strongly coupled at the cutoff. One can consider the uncertainty plot of \cite{Dimopoulos:2002mv} to essentially apply to our setup as well.
If we require the observed four-dimensional gauge coupling $g_4^2$ is order one, then $g_5^2/ \pi R \sim 1$ and hence $\delta_i \sim (\Lambda \pi R)^{-1}$. The result is simply that the larger the volume, the smaller the theoretical uncertainty.

\subsection{Running}
We can now consider quantum effects. Because the spectrum of the theory, 
including Kaluza-Klein (KK) modes, is not $SU(3)$ symmetric, we will generate 
log enhanced contributions to the $\delta_i$ in eq. (\ref{error}). 
These are completely calculable within an effective theory. One 
straightforward approach is to sum the contributions of the 
KK modes along the lines of \cite{Dienes:1998vg,Dienes:1998vh}. We can then 
calculate the effective four-dimensional couplings as a function of energy
\bea
\alpha^{-1}(\mu)& =& \alpha^{-1}(\mu_0)-\frac{b}{2 \pi} \log(\mu/\mu_0)\\ &-&
\frac{{\tilde b_{even}}}{8 \pi} r_{even}(\mu,R) 
-\frac{{\tilde b_{odd}}}{8 \pi} r_{odd}(\mu,R).
\nonumber
\eea
We define
\bea
&r_{even}(\mu,R) &=  \int_{\pi/ 4 \mu^2}^{\pi R^2/4} \>dt
\frac{\theta_3(i t/\pi R^2)-1}{t}, \\
&r_{odd}(\mu,R)&= \nonumber
\\ & &\int_{\pi/ 4 \mu^2}^{\pi R^2}\>dt 
\frac{\theta_3(i t/\pi 4 R^2)-\theta_3(i t/\pi R^2)}{t}, \hskip 0.3in
\nonumber
\eea
where $\theta_3(t)=\sum_{n=-\infty}^{\infty} \exp(\pi i t n^2)$. Here, $\tilde b_{even}$ and $\tilde b_{odd}$ are the contributions to the beta functions from the modes at $n/R$ and $(2n+1)/2 R$, respectively.

While the above expression actually encodes the power law running due to the KK tower, the overwhelming majority will be $SU(3)$ universal, with only a logarithmic piece distinguishing $\alpha_Y$ from $\alpha_2$. This can be studied by looking at the relative running of the two couplings (suitably normalized)
\bea
\delta \alpha^{-1}(\mu)&=& \alpha_Y^{-1}(\mu)-3 \alpha_2^{-1}(\mu) 
\nonumber \\
&=& \alpha^{-1}_Y(m_z)-3\alpha_2^{-1}(m_z)\\ \nonumber
&-&\frac{b_Y-3 b_2}{2 \pi}\log(\mu/\mu_0)\\ \nonumber
&-&\frac{\tilde b_{Y,even}-3 \tilde b_{2,even}}{8 \pi} r_{even}(\mu,R)\\
\nonumber
&-&\frac{\tilde b_{Y,odd}-3 \tilde b_{2,odd}}{8 \pi} r_{odd}(\mu,R).
\eea
This quantity encodes the relative running between the two couplings, and by studying where it crosses zero, we can determine the cutoff of the theory.

\section{Non-supersymmetric models}
In this section we present non-supersymmetric theories of 
$SU(3)$ electroweak unification in five dimensions.  We describe
two models explicitly, first a simple extra-dimensional version
of the minimal module \cite{Dimopoulos:2002mv} and then a version in which
quarks and leptons are on different boundaries, with the Higgs
in the bulk, thus naturally explaining the absence of proton decay 
and explaining why leptons appear to come in complete $SU(3)$
triplets.

In the theories below there are two theoretical inputs, $\Lambda$
and $R$, to which the value of $\sint$ is logarithmically sensitive.
The range of these parameters are restricted by the hierarchy problem.
If the theory is not strongly coupled at $\Lambda$ we expect one-loop 
corrections to the squared Higgs mass to be of order a loop factor
times $\Lambda^2$.  To produce a Higgs mass around the electroweak
breaking scale without more than 1\% fine tuning, we require 
$\Lambda< 20$ TeV.  The value of $1/R$ is then restricted to the
range 1 TeV $\leq 1/R < \Lambda$.  For the cutoff in the range
2--20 TeV, the variation of $\sint$ at $M_z$ is roughly a few percent.

~From the previous section, another contribution to the uncertainty
are boundary kinetic terms.  A guess at the fractional uncertainty
in the squared couplings is $\sim 1/(\pi R \Lambda)$ which can be somewhat
large in the scenarios outlined below.  However, for large regions
in parameter space \cite{Dimopoulos:2002mv}, the corrections are in fact quite small,
again of order a few percent.

\subsection{Minimal Reconstruction}
This model is simply a continuous version of the two-site moose
shown in Figure \ref{drcon}.  An $SU(3)$ gauge theory lives in the
full five dimensions but is broken by orbifold boundary conditions
at $y=\pi R$ by requiring the following properties of the gauge fields:
\bea
A^{\mu}(- y) &=& A^{\mu}(y) \\ \nonumber
A^{\mu}(2\pi R - y) &=& Z A^{\mu}(y) Z
\eea
and
\bea
A^{5}(- y) &=& - A^{5}(y) \\ \nonumber
A^{5}(2\pi R - y) &=& - Z A^{5}(y) Z
\eea
where
\be
Z = \pmatrix{-1&0&0 \cr 0&-1&0 \cr 0&0&\phantom{-}1}\, ,
\label{eq:z}
\ee
breaks $SU(3)\rightarrow SU(2)\times U(1)$.  The massive gauge bosons
form an $SU(2)$ doublet with hypercharge 3/2.  The masses of states
in the KK towers of the unbroken and broken generators are $n/R$
and $(n+1/2)/R$ respectively.  The differential running of gauge 
couplings can be calculated using the techniques outlined in the 
previous section.  For this model the beta function coefficients
for the zero modes are $(b,b_Y)=(-19/6,41/6)$, and for the KK modes are 
$(\tilde b_{2,even}, \tilde b_{Y,even}; \tilde b_{2,odd}, \tilde b_{2,odd})
= (-21/3,0; -21/6,-63/2)$.  

We can now compute the size of the extra dimension as a function of the
cutoff and find, for example, for $\Lambda = (10,20)$ TeV we get 
$L^{-1}\equiv (\pi R)^{-1} \sim (1.9,1.3)$ TeV.  However, a tiny difference
in the gauge couplings (percent) results in a significant difference in 
the compactification scale (order one).  The sensitivity of $1/R$ to the exact
gauge couplings translates into an insensitivity of $\sint$ to the specific
values of $1/R$ chosen above. 

\begin{figure}
\centerline{
\psfig{file=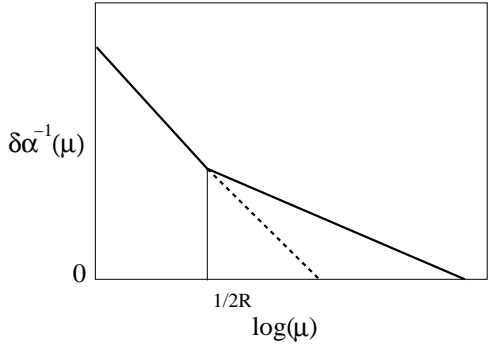,width=0.4\textwidth}}
\caption{Differential running of the $SU(2)$ and $U(1)$ gauge couplings.
	The dashed line represents the running in the absence of an
	extra dimension.}
\label{fig:running}
\end{figure}

Variations of this model involve moving the Higgs and leptons off of the
$SU(2)\times U(1)$ boundary.  This is possible because they carry quantum
numbers of components of $SU(3)$ representations while the quarks do not.
(Putting leptons in the bulk does, however, require doubling the number of
species as the orbifold projections only allow one component of each triplet
to survive.)  The renormalization group analysis will differ but produces
the same generic features of the model above as displayed in 
Figure \ref{fig:running}, namely that the relative running is slowed by the 
existence of partial multiplets in the KK towers.

\subsection{Proton Stability}
A particularly intriguing model is motivated by placing the SM fields
in their ``natural'' locations based on their quantum numbers.  The quarks
must remain on the $SU(2)\times U(1)$ while the leptons fall into
{\it complete} $SU(3)$ multiplets \cite{Konopinski:1953gq} 
and thus should live on 
the $SU(3)$ preserving boundary.  The Higgs 
does not fill out a complete multiplet and 
thus should live in the bulk where the orbifolding splits multiplets.  
The Higgs could live in a {\bf 3} or ${\bf {\bar 6}}$ of $SU(3)$.  The 
former would only allow the highly constrained lepton Yukawa couplings 
$y_{\ell}^{ij} H L^i L^j$ contracted with an epsilon tensor predicting 
$m_{\mu} = m_{\tau}$ and $m_e = 0$.  The ${\bf {\bar 6}}$ however is a 
symmetric tensor and easily allows enough freedom to produce the charged 
lepton spectrum.  The Higgs in the bulk will satisfy
\bea
H(- y) &=& H(y) \\ \nonumber
H(2\pi R - y) &=& - Z H(y) Z
\eea
where $Z$ is defined as in equation \ref{eq:z}.

thus projecting out an $SU(2)$ triplet with hypercharge -1 and a singlet
with hypercharge +2 and leaving only the zero mode for the charge 1/2
doublet.  The zero mode spectrum is just the standard model and the Higgs
is expected to get a mass at one loop due predominantly to the top 
Yukawa coupling, self coupling and gauge couplings.

The other attractive component of this scenario is that it naturally
suppresses proton decay as no local counter terms containing both quarks
and leptons can be constructed \cite{Arkani-Hamed:1999dc}. However, 
one can imagine exponentially suppressed
contributions through the exchange of heavy states.  For a cutoff
of 20 TeV, an extra dimension of size $\sim 1$ TeV, and weak coupling at 
the cutoff, dimension 6 operators are sufficiently suppressed if the 
relevant states to be exchanged are at least a few times the cutoff.

This model is particularly interesting as it requires (by gauge
invariance) a coupling of the leptons to the exotic singly and 
doubly charged gauge bosons.  If produced, the doubly charged
gauge bosons will decay into like-sign leptons with very high $p_T$,
something easily seen at RUN II or the LHC, depending on its mass.

\section{Supersymmetric Models}
While these are elegant models in which $\sint$ is predicted, we have only addressed questions of the hierarchy problem in the sense that the cutoff of the theory is low. Moreover, we have no understanding of the origin of electroweak symmetry breaking. To this end, there has been great interest in supersymmetric models in TeV-sized extra dimensions.

Theories with TeV-sized dimensions provide the simplest framework for explaining electroweak symmetry breaking. This can be realized either with the Scherk-Schwarz mechanism, or with a large, localized supersymmetry breaking term, both of which have the added benefit of solving the supersymmetric flavor problem automatically. These mechanisms naturally relate the scale of EWSB to the size of the extra dimension. In these scenarios, gauginos naturally have mass of $1/R$, while the sfermions are generated at one loop order higher (i.e., $~(4 \pi R)^{-1}$). 

There is still the free parameter $\Lambda$, which is limited by strong coupling to be within a few orders of magnitude of $R^{-1}$. 
If one makes the additional assumption that $\Lambda$ is the strong coupling scale of the theory, one can then make a more robust prediction of $\sint$.

In addition to the location of the Higgs and 
leptons, one has other possibilities for constructing models. Supersymmetry can be broken by 
Scherk-Schwarz boundary conditions, such as in \cite{Antoniadis:1990ew,Antoniadis:1998sd,Delgado:1998qr}.\footnote{The consistency of the 
continuous Scherk-Schwarz breaking with orbifold breaking of gauge groups 
has been explored in \cite{Barbieri:2001yz}.} One can also give a large boundary mass 
for the gauginos, but this limits the setup to having all matter and 
Higgs fields on the $SU(2)\times U(1)$ boundary in order to avoid flavor 
violation and/or very large ($O(R^{-1})$) soft masses for the Higgses.

\begin{figure}
\centerline{
\psfig{file=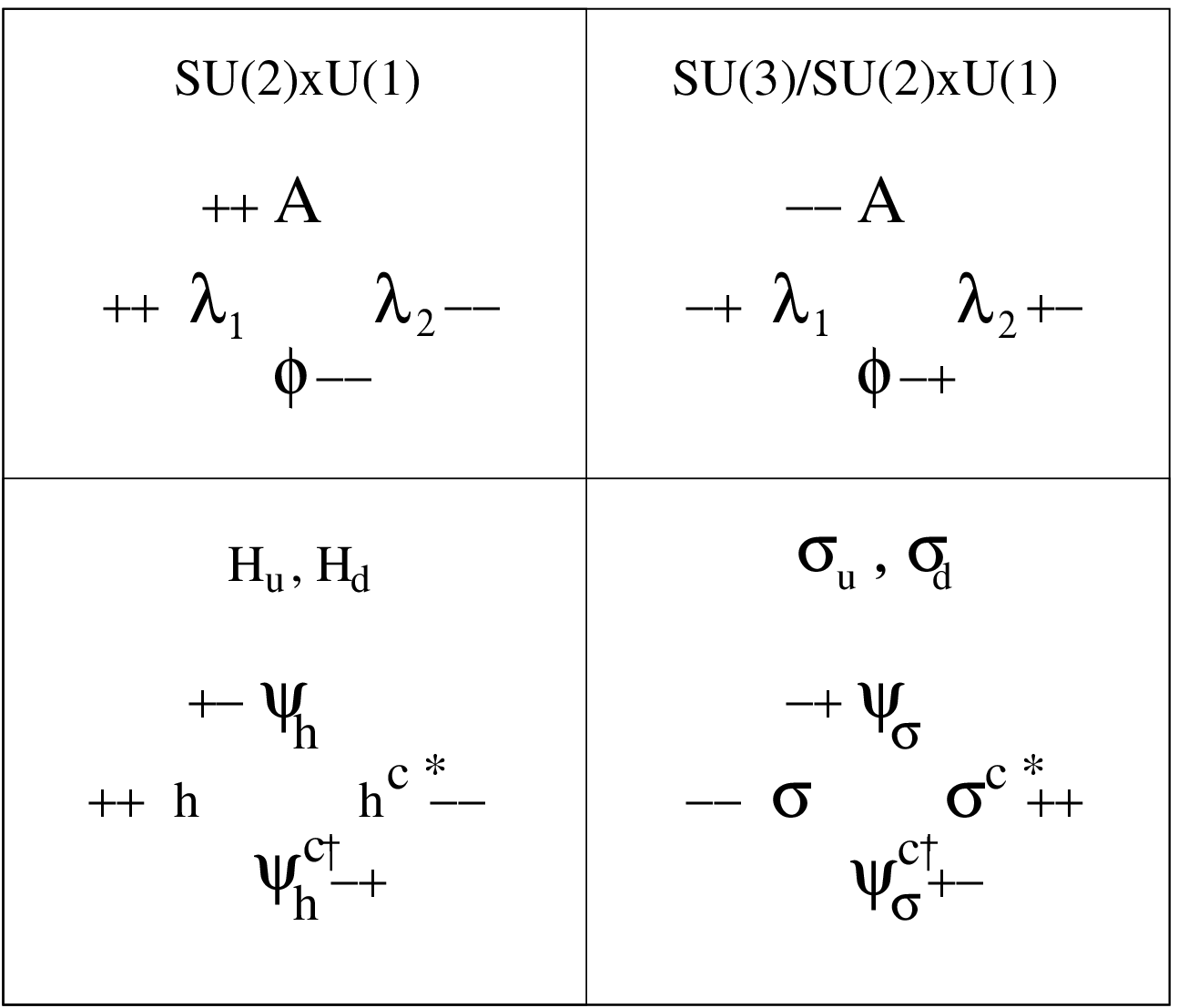,width=0.4\textwidth}}
\caption{Charge assignments for a model with SU(3) broken to SU(2)$\times$U(1) on both boundaries.}
\label{fig:charges}
\end{figure}

Such models have a few phenomenologically interesting features: first, with Scherk-Schwarz compactifications and gaugino masses on the $SU(3)$ boundary, it is quite natural to have binos which are nearly degenerate with the winos. If the leptons live in a triplet on the $SU(3)$ boundary (with a sextet Higgs in the bulk, as described), the right-handed sleptons will receive additional gauge mediated contributions from the broken generators, potentially reflecting the underlying $SU(3)$ symmetry. Interestingly, with quarks and leptons on separated boundaries, we need only ordinary R-parity to forbid dangerous proton decay operators. (Without R-parity, superpotential couplings like $QQQ(H_d+\partial_y H_u^c)$ and $L(H_u+\partial_y H_d^c)$ together can lead to unacceptably large proton decay rates.)

Unification is not a generic feature of these models.  
However, as an example of an interesting model in which unification does occur, consider the following scenario: we will assume both boundaries to be broken to $SU(2)\times U(1)$, with quarks on one boundary, and the leptons on the other. Higgses and gauge fields propagate in the bulk with the charge assignments of figure \ref{fig:charges}. In this model, all KK modes come in complete $SU(3)$ multiplets, so only the zero modes contribute. The zero mode beta functions are $(b'=25/2,b_2=-5/6)$. In addition to sfermions which will contribute to the running at a scale $\sim (4 \pi R)^{-1}$, there is also the scalar singlet under $SU(2)$ with hypercharge 1, as well as the fifth component of the broken generator gauge bosons which transforms as a doublet under $SU(2)$ with hypercharge $3/2$. These particles should also pick up masses down by a loop factor from $R^{-1}$. With this set up, and a compactification scale of 1 TeV, we find unificaiton at 5.3 TeV.

This model illustrates again the possibility of new, doubly charged particles at the weak scale. The decay modes are model dependent, but generically one would expect again hard charged leptons. The $SU(2)$ singlet field would most likely decay through an off-shell broken gauge boson into a Higgs. The decays of these would then give hard leptons and b quarks (or W bosons if the Higgs is sufficiently heavy, which can happen in TeV scale 5D SUSY theories \cite{Arkani-Hamed:2001mi,Weiner:2001ui}).

Ultimately, there is a wealth of phenomenology to be explored, but aside from these few generic points, it is best studied within the context of a complete model in which EWSB is realized.

\noindent{\bf Conclusions:} In this paper we used the mechanism of ref 
\cite{Dimopoulos:2002mv} to reconstruct TeV-scale
five-dimensional theories that successfully predict the weak angle. The 
best limit to the size of the new dimensions
comes from the mass limits of excited $Z$s and is $1/R>1.3 TeV$. There is a 
plethora of experimental predictions, such
as the presence of exotic singly and doubly charged gauge bosons. Direct 
searches place a limit of 700 GeV to their
mass. They can decay into pairs of (same sign)leptons. Even the lightest 
new particles will in general decay through
higher dimension operators to particles on the boundaries and may give 
dramatic signatures at the LHC.


\noindent While completing this work, we became aware of refs \cite{Li:2002pb,Hall:2002rk}, which address similar issues.

{\bf Acknowledgments:} The authors would like to thank N.Arkani-Hamed for useful discussions. This work is supported by NSF grant PHY-9870115 and DOE grant DE-AC0376SF00515. NW was supported by DOE grant DE-FG03-96-ER40956.

\bibliography{orb2} 
 
\bibliographystyle{JHEP}









\end{document}